\documentclass[12pt]{article}
\usepackage{amsfonts, latexsym}

\newtheorem{lemma}{Lemma}

\begin{document}


\title{\bf On the Gauss Law and Global Charge for
QCD}

\author{
    J. Kijowski \\
    Center for Theoretical  Physics, Polish Academy of Sciences\\
    al. Lotnik\'ow 32/46, 02-668 Warsaw, Poland\\
    \ \\
    G. Rudolph \\
    Institut f\"ur Theoretische Physik, Universit\"at Leipzig\\
    Augustusplatz 10/11, 04109 Leipzig, Germany\\
    }

\maketitle


\begin{abstract}
The local Gauss law of quantum chromodynamics on a finite lattice
is investigated. It is shown that it implies a gauge invariant,
additive law giving rise to a gauge invariant ${\mathbb
Z}_3$-valued global charge in QCD. The total charge contained in a
region of the lattice is equal to the flux through its boundary of
a certain ${\mathbb Z}_3$-valued, additive quantity. Implications
for continuous QCD are discussed.
\end{abstract}

\newpage

\vspace{0.5cm}



\setcounter{equation}{0}
\section{Introduction}
Quantum chromodynamics (QCD) is one of the basic building blocks
of the standard model for describing elementary particle
interactions. During the last decades QCD was quite successful,
e.g. in describing deep inelastic scattering processes within the
framework of perturbation theory at one hand and ``measuring''
certain types of observables using (nonperturbative) lattice
approximation techniques on the other hand. Nonetheless, we are
still facing the basic challenge, which consists in constructing
an effective microscopic theory of interacting hadrons out of this
gauge theory. For this purpose, nonperturbative methods for
describing the low energy regime should be developed. In
particular, the observable algebra and the superselection
structure of this theory should be investigated. The present paper
is a step in this direction.

We stress that standard methods from algebraic quantum field
theory for models which do not contain massless particles, see
\cite{DHR}, do not apply here. Some progress towards an
implementation of similar ideas for theories with massless
particles has been made, for the case of quantum electrodynamics
(QED) see \cite{JF} and \cite{Bu} and further references therein.
In particular, Buchholz developed the concept of so called charge
classes and found a criterion for distinguishing between the
electric charge (carried by massive particles) and additional
superselection sectors corresponding to inequivalent asymptotic
infrared clouds of photons. For some attempts to deal with the
nonabelian case we refer to papers by Strocchi and Wightman, see
\cite{SW1} and \cite{S}.

In QED, the notion of {\em global} (electric) charge is easy to
understand. This is due to the fact, that in this theory we have a
{\em local} Gauss law, which is built from gauge invariant
operators and which is linear. Thus, one can ``sum up'' the local
Gauss laws over all points of a given (spacelike) hyperplane in
space time yielding the following gauge invariant conservation
law: The global electric charge is equal to the electric flux
through a $2$-sphere at infinity. On the contrary, in QCD the
local Gauss law is neither built from gauge invariant operators
nor is it linear. The main point of the present paper is to show
that it is possible to extract from the local Gauss equation of
QCD a gauge invariant, additive law for operators with eigenvalues
in ${\mathbb Z}_3$, the center of $SU(3)$. This implies -- as in
QED -- a gauge invariant conservation law: The global ${\mathbb
Z}_3$-valued charge is equal to a ${\mathbb Z}_3$-valued gauge
invariant quantity obtained from the color electric flux at
infinity. There is a paper by Fredenhagen and Marcu, see
\cite{FM}, dealing with a ${\mathbb Z}_2$-gauge theory (with
${\mathbb Z}_2$-valued matter fields) on the lattice. These
authors were able to construct the ground state and charged states
of this model. For some regions in the space of coupling
constants, the thermodynamic limit of charged states was
controlled. It is likely that similar techniques will be relevant
for studying ${\mathbb Z}_3$-valued colour charged states of QCD.

We stress that our main observation is universal in the following
sense: Suppose somebody had constructed a nonperturbative version
of QCD rigorously. Then our construction of the global charge
would apply. Since, however, such a theory is not at our disposal,
for a rigorous discussion we have to restrict ourselves to the
case of QCD approximated on a finite lattice. For basic notions
concerning lattice gauge theories (including fermions) we refer to
\cite{Seiler} and references therein. Thus, we consider QCD  on a
finite, regular cubic lattice in the Hamiltonian approach. Our
starting point is the notion of the algebra of field operators.
Since, up to our knowledge, this has never been published, we
discuss it in some detail. By imposing the local Gauss law and
gauge invariance we obtain the observable algebra ${\cal
O}(\Lambda )$ for QCD on the lattice. For an analogous
investigation of the superselection structure of QED on the
lattice we refer to \cite{KRT} and for a (rather heuristic)
discussion of continuum QCD within the functional integral
approach (using only gauge invariant quantities) we refer to
\cite{KRR}. For QED, the mathematical structure of ${\cal
O}(\Lambda )$ has been completely clarified, see \cite{KRS}. A
similar investigation for QCD seems to be a very difficult
problem, which will be addressed in the future. In this paper, we
concentrate on a careful analysis of the local Gauss law and its
consequences for the notion of global charge in QCD.

Our paper is organized as follows: In Sections \ref{Algebra} and
\ref{Observablealgebra} we discuss the algebra of fields and the
observable algebra for QCD on a finite lattice. In Section
\ref{Gauss Law} we analyse the local Gauss law and invent the
gauge invariant notion of the local charge densitiy. Using this
notion, in Section \ref{Charge} the global charge is defined and
the flux law for QCD is discussed. Here also some heuristic
remarks concerning the continuum case are added.


\setcounter{equation}{0}
\section{The Field Algebra for Lattice QCD}
\label{Algebra}


We consider QCD in the Hamiltonian framework on a finite, regular
3-dimensional lattice $\Lambda$. We denote the set of oriented,
$i$-dimensional elements of $\Lambda$ by ${\Lambda}^i$, $i =
0,1,2,3 \, .$ Such elements are (in increasing order of i) called
sites, links, plaquettes and cubes. The set of non-oriented
$i$-dimensional elements will be denoted by $|{\Lambda}|^i$. If,
for instance, $(x,y) \in {\Lambda}^1$  is an oriented link, then
by $|(x,y)| \in |{\Lambda}|^1$ we mean the corresponding
non-oriented link. Instead of using a concrete Hilbert space
representation (e.g. the Schr\"odinger representation), we give an
abstract definition of the field algebra in terms of generators
and defining relations.

The basic fields of lattice QCD are quarks living at lattice sites
and gluons living on links. The field algebra is thus, by
definition, the tensor product of local fermionic and bosonic
algebras:
\begin{equation}
\label{fieldalgebra} {\cal F}(\Lambda) :=  \bigotimes_{x \in
|\Lambda |^0} {\cal F}_x \bigotimes_{|(x,y)| \in |\Lambda |^1}
{\cal F}_{|(x,y)|} \, .
\end{equation}
We impose locality of the lattice quantum fields by postulating
that the local algebras corresponding to different elements of
$\Lambda$ commute with each other.

The fermionic field algebra ${\cal F}_x$ associated with a lattice
site $x$ is the enveloping algebra generated from the complex Lie
super-algebra ${\cal L}_x$ of canonical anticommutation relations
of quarks. In terms of coordinates, the quark field is given by
\begin{equation}
|{\Lambda}|^0 \ni x \rightarrow  ({\psi}^{aA}(x)) \in {\cal L}_x
\, ,
\end{equation}
where $a$ stands for bispinorial and (possibly) flavour degrees of
freedom and $A = 1,2,3$ is the colour index corresponding to the
fundamental representation of $SU(3)$. The conjugate quark field
is denoted by ${\psi^{*a}}_{A}(x)  \, ,$ where we raise and lower
indices by the help of the canonical hermitian metric tensor
$g_{AB}$ in $\mathbb C^3$. Finally, the (nontrivial) canonical
anticommutation relations for elements of ${\cal L}_x$ read:
\begin{equation}
\label{CCR3} [{\psi^{*a}}_{A}(x), \psi^{bB}(x)]_+ = {\delta^B}_A
\delta^{ab} \, .
\end{equation}
Passing to self-adjoint generators (real and imaginary parts of
$\psi$), we observe that ${\cal F}_x$ is a Clifford algebra.

The bosonic field algebra ${\cal F}_{|(x,y)|}$ associated with the
non-oriented link $|(x,y)|$ will be constructed in terms of its
equivalent copies ${\cal F}_{(x,y)}$ and ${\cal F}_{(y,x)}$,
corresponding to the two orientations of the link $(x,y)\, .$ We
will see that there is a natural identification of these two
algebras, induced from the vector character of both the gluonic
potential $A$ and the colour electric field $E$ of the underlying
continuum theory: under the change of orientation of the $k$'th
axis both $A_k$ and $E^k$ change their signs.

The bosonic field algebra ${\cal F}_{(x,y)}$ is, by definition,
the tensor product
\begin{equation}
\label{balgebra-xy} {\cal F}_{(x,y)} :=  \tilde{\cal E}_{(x,y)}
\otimes {\cal A}_{(x,y)} \, ,
\end{equation}
where $\tilde{\cal E}_{(x,y)}$ is the enveloping algebra of the
(real) Lie algebra ${\cal E}_{(x,y)} \cong su(3)$ and ${\cal
A}_{(x,y)} \cong C^\infty(SU(3))$ is the commutative $*$-algebra
of smooth functions on the Lie group $SU(3) \, .$ We note that the
above tensor product is naturally endowed with the structure of a
crossed product of Hopf algebras, given by the action of
generators of $\tilde{\cal E}_{(x,y)}$ on functions, see
\cite{KS}. We identify the tensor product of elements of
$\tilde{\cal E}_{(x,y)}$ and functions with their product as
differential operators on $SU(3) \, .$ This way, ${\cal
F}_{(x,y)}$ gets identified with the algebra of differential
operators on the group manifold. For our purposes, we endow ${\cal
F}_{(x,y)}$ with a Lie algebra structure. Thus, we have to define
the commutator between a generator $e \in {\cal E}_{(x,y)}$ of
$\tilde{\cal E}_{(x,y)}$ and a function $f \in {\cal A}_{(x,y)} \,
,$
\begin{equation}
\label{commut-e-f} [e , f ] := e^R(f)  \ ,
\end{equation}
where $e^R(f)$ denotes the derivative of $f$ with respect to the
right-invariant vector field $e^R \, $ generated by $e \, ,$
\begin{equation}\label{derivative}
e^R(f)(g) := \frac d{ds}{\Big |}_{s=0} f(\exp (se)\cdot g) \, , \,
\, g \in SU(3) \, .
\end{equation}

Now we give an explicit description of ${\cal F}_{(x,y)}$ in terms
of generators and defining relations. The algebra  ${\cal
A}_{(x,y)}$ is generated by matrix elements of the gluonic gauge
potential on the link $(x,y) \, ,$
\begin{equation}
{\Lambda}^1 \ni (x,y) \rightarrow  {U^A}_B(x,y) \in  {\cal
A}_{(x,y)} \, ,
\end{equation}
where $A,B = 1,2,3$ are colour indices. Being functions on $SU(3)
\, ,$ these generators have to fulfil the following relations:
\begin{eqnarray}
\label{unitary1} ({U^A}_B(x,y))^* {U_A}^C(x,y) & = & {\delta^C}_B
\ {\bf 1} \ ,
\\
\epsilon_{ABC}\  {U^A}_D(x,y) {U^B}_E(x,y) {U^C}_F(x,y) & = & \
\epsilon_{DEF} \ {\bf 1} \ . \label{unitary2}
\end{eqnarray}

The algebra $\tilde{\cal E}_{(x,y)}$ is generated by colour
electric fields, spanning the Lie algebra ${\cal E}_{(x,y)}$.
Choosing a basis $\{t_i \} \, ,$ $i = 1, \dots , 8$, of $su(3)$ we
denote by $\{ E_i(x,y)\}$ the corresponding basis of ${\cal
E}_{(x,y)}$,
\begin{equation}
{\Lambda}^1 \ni (x,y) \rightarrow  {E}_i(x,y) :=  t_i \in {\cal
E}_{(x,y)} \, .
\end{equation}
These generators are self-adjoint (real), $E_i^* = E_i \, .$ In
the sequel we take as the basis the hermitean, traceless Gell-Mann
matrices ${{t_i}^A}_B \, ,$ normalized as follows:
\begin{equation}
\sum_i { { t_i}^A}_B { { t_i}^C}_D = \delta^A_D \delta^C_B - \frac
13 \delta^A_B \delta^C_D \, ,
\end{equation}
or, equivalently, $  { { t_i}^A}_B \ { { t_j}^B}_A = \delta_{ij}
$. We will also use the following traceless matrix built from
these fields
\begin{equation}
{E^A}_B(x,y) = \sum_i E_i (x,y) { { t_i}^A}_B \  . \label{EABEi}
\end{equation}
Since the coefficients ${ { t_i}^A}_B$ are complex, these fields
are no longer self-adjoint:
\begin{eqnarray}
\label{self-adj} ({E^A}_B(x,y))^* & = & {E_B}^A(x,y) \, .
\end{eqnarray}
The commutation relations between elements of $su(3)$ translated
to the language of these fields read
\begin{equation}
\label{CCR2} [{E^A}_B(x,y) , {E^C}_D(x,y)] = {\delta^C}_B
{E^A}_D(x,y)  - {\delta^A}_D {E^C}_B(x,y)   \ ,
\end{equation}
whereas the commutation relations (\ref{commut-e-f}) between Lie
algebra elements and functions, rewritten in terms of generators,
take the following form:
\begin{equation}
\label{CCR1} [{E^A}_B(x,y) , {U^C}_D(x,y)] = {\delta^C}_B
{U^A}_D(x,y) - \frac{1}{3} {\delta^A}_B {U^C}_D(x,y)   \, .
\end{equation}

Observe that, for every link $(x,y)\, ,$ we have a model of
quantum mechanics with configuration space being the group
manifold $SU(3)\, .$ The matrix elements of the gluonic potential
play the role of functions of position variables, whereas the
colour electric fields play the role of (non-commuting)
canonically conjugate momenta. Formula (\ref{CCR1}) is the analog
of the canonical commutation relation $[p,q]= - i \, .$

\vspace{0.2cm} \noindent Now, we describe the transformation law
of these objects under the change of the link orientation. This
law is an isomorphism between bosonic field algebras:
\begin{equation}
\label{izo-I} {\cal I}_{(x,y)} : {\cal F}_{(x,y)} \rightarrow
{\cal F}_{(y,x)} \, .
\end{equation}
The vector character of the continuum gauge potential implies that
the (classical) $SU(3)$-valued parallel transporter $g(x,y)$ on
$(x,y)$ transforms to $g^{-1}(x,y)$ under the change of
orientation. Thus, the change of orientation on $(x,y)$ induces
the transformation
\begin{equation}
\label{inverse} SU(3) \ni g \rightarrow i(g) = g^{-1} \in SU(3)
\end{equation}
on configuration space. This transformation lifts naturally to the
field algebra ${\cal F}_{(x,y)} \, .$ Indeed, for elements of
${\cal A}_{(x,y)}$ it implies the following transformation law
\begin{equation}
\label{izo-f} {\cal A}_{(x,y)} \ni f \rightarrow  \breve{f}  \in
{\cal A}_{(y,x)} \, ,
\end{equation}
where $\breve{f}(g) := f(g^{-1}) \, .$ We thus put ${\cal
I}_{(x,y)}(0,f) := (0, \breve{f} )$. It may be easily proved that
the unique $*$-isomorphism, which fulfils this requirement is
given by
\begin{equation}
\label{izo-I1} {\cal I}_{(x,y)}(e,f) := (e^L, \breve{f} ) \ ,
\end{equation}
where $e^L$ is the left invariant vector field on $SU(3)$
generated by $-e$. It acts on functions in the following way:
\begin{equation}\label{l-derivative}
    e^L(f)(g) = \frac d{ds}{\Big |}_{s=0} f(g \cdot \exp (-se)) \ .
\end{equation}
Observe that $e^L$ is not an element of ${\cal E}_{(y,x)}$, but it
can be expanded with respect to right-invariant vector fields with
coefficients being functions on the group. Actually, the following
formula is easily proved:
\begin{equation}\label{L-R}
  e^L = - \sum_{i=1}^8 \mbox{\rm Tr}(g e g^{-1} t_i ) t^R_i \ ,
\end{equation}
where $t_i$ is any orthonormal basis of $su(3)$.

To prove that ${\cal I}_{(x,y)}$ is an isomorphism indeed, we
observe that it is obviously a bijective mapping between
generators. It extends to an isomorphism of field algebras if we
prove that it preserves the commutator. This is obvious for
elements, which either belong both to ${\cal A}_{(x,y)}$ or to
${\cal E}_{(x,y)}$. Thus, it is sufficient to consider the
commutator of a Lie algebra element $e$ with a function $f\, .$
Using the identity
\begin{equation}
\label{odwracanie}
  e^R(f)(g) = \frac d{ds}{\Big |}_{s=0} f(\exp (se)\cdot g) =
    \frac d{ds}{\Big |}_{s=0} \breve{f}(g^{-1} \cdot \exp (-se)) =
    e^L(\breve{f})(g^{-1}) \ ,
\end{equation}
and applying ${\cal I}_{(x,y)}$ to the functions on both sides, we
obtain due to (\ref{commut-e-f}) and (\ref{izo-f}):
\begin{equation}
\label{izo-ef}
  {\cal I}_{(x,y)} \left( [ e , f ] \right) = {\cal
  I}_{(x,y)}(e^R (f)) = e^L(\breve{f}) = \left[
  {\cal I}_{(x,y)}(e) , {\cal I}_{(x,y)}(f) \right] \ .
\end{equation}
This shows that ${\cal I}_{(x,y)}$ is an isomorphism, indeed. We
add that, in more abstract terms, it is given by $ {\cal
I}_{(x,y)}(e,f) = (-i_*(e),i^*(f) ) \, .$

Field configurations, which are related under ${\cal I}_{(x,y)}$
will be identified as different representations of the same
object. Thus, the bosonic field algebra ${\cal F}_{|(x,y)|} \, ,$
associated with the non-oriented link $|(x,y)|\, ,$ is defined as
the subalgebra of ${\cal F}_{(x,y)} \times {\cal F}_{(y,x)} \, ,$
obtained by this identification:
\begin{eqnarray}
\label{gluons}
  {\cal F}_{|(x,y)|}&:=& \left\{ (l_{(x,y)} , k_{(y,x)} ) \in
{\cal F}_{(x,y)} \times {\cal F}_{(y,x)} \, : \,
  k_{(y,x)} = {\cal I}_{(x,y)} (l_{(x,y)}) \, \right\}
\ .
\end{eqnarray}
Projection onto the first (resp.~second) component gives us an
isomorphism of ${\cal F}_{|(x,y)|}$ with ${\cal F}_{(x,y)}$ (resp.
${\cal F}_{(y,x)}$).

The above transformation law yields relations between generators of
the two algebras. For functions on $SU(3)$, formula
(\ref{unitary1}) enables us to rewrite (\ref{izo-f}) as follows:
\begin{equation}
\label{izo-U}
  {U^A}_B(y,x) = ({U_B}^A(x,y))^* \ .
\end{equation}
For Lie algebra elements, formula (\ref{L-R}) applied to $e = t_i$
reads:
\begin{equation}
\label{er-el}
    {E}_j(y,x) = - \sum_{i=1}^8 {U^A}_B(x,y) {{t_j}^B}_C {U^C}_D(y,x)
{{t_i}^D}_A \  {E}_i(x,y) \, ,
\end{equation}
or, in terms of generators ${E^A}_B(x,y) \, ,$
\begin{eqnarray}\label{ER-EL-1}
   {E^A}_B(y,x) & = & - {U^A}_D(y,x) {U^C}_B(x,y) {E^D}_C(x,y)  \ .
\end{eqnarray}

\vspace{0.2cm} \noindent To summarize, the field algebra ${\cal
F}(\Lambda)$ is the $*$-algebra, generated by elements
\[
\left\{ \psi^{aA}(x) \, , \, {U^A}_B(x,y) \, , \, {E^A}_B(x,y)
\right\} \, ,
\]
fulfilling relations (\ref{unitary1}), (\ref{unitary2}),
(\ref{self-adj}), (\ref{izo-U}) and (\ref{ER-EL-1}), together with
canonical (anti-) commutation relations
\begin{eqnarray}
 [{\psi^{*a}}_{A}(x), \psi^{bB}(y)]_+ & = & \delta(x-y)
{\delta^B}_A \delta^{ab} \, , \nonumber\\ \left[{E^A}_B(x,y) ,
{E^C}_D(u,z)\right] & = & \delta(x-u) \delta(y-z)
\left({\delta^C}_B {E^A}_D(x,y)  - {\delta^A}_D {E^C}_B(x,y)
\right)  \, , \nonumber\\ \left[{E^A}_B(x,y) , {U^C}_D(u,z)
\right] & = &  + \, \,  \delta(x-u) \delta(y-z) \left({\delta^C}_B
{U^A}_D(x,y) -\frac{1}{3} {\delta^A}_B {U^C}_D(x,y) \right)
\nonumber \\ & & - \, \, \delta(x-z) \delta(y-u)
\left({\delta^A}_D {U^C}_B(y,x) -\frac{1}{3} {\delta^A}_B
{U^C}_D(y,x) \right)  \, . \nonumber
\end{eqnarray}

\section{The Observable Algebra}
\label{Observablealgebra}

The observable algebra ${\cal O}(\Lambda )$ is defined by imposing
the local Gauss law and gauge invariance.

The group of local gauge transformations acts on ${\cal F}(\Lambda
)$ by automorphisms $x \rightarrow {g^A}_B(x) $ as follows:
\begin{eqnarray}
\label{gauge} \psi^{aA}(x) & \rightarrow & {g^A}_B(x) \psi^{aB}(x)
\ ,
\\
{U^A}_B(x,y) & \rightarrow & {g^A}_C(x) {U^C}_D(x,y)
{{(g^{-1})}^D}_B(y) \ ,
\\
{E^A}_B(x,y) & \rightarrow & {g^A}_C(x) {E^C}_D(x,y)
{{(g^{-1})}^D}_B(x) \ .
\end{eqnarray}
It is easy to check that these transformations are generated by
\begin{equation}
\label{calG}
  {{\cal G}^A}_B(x) := {\rho^A}_B(x) - \sum_y {E^A}_B(x,y)  \,,
\end{equation}
where
\begin{equation}\label{rho}
{\rho^A}_B(x) = \sum_a \left( \psi^{*aA}(x) {\psi^a}_{B}(x) -
\frac 13 \delta ^A_B \psi^{*aC}(x) {\psi^a}_{C}(x) \right) \
\end{equation}
is the local matter charge density. Observe that ${\rho^A}_A(x) =
0 \, .$

To implement gauge invariance we have to take those elements of
${\cal F}(\Lambda )\, ,$ which commute with all generators $
{{\cal G}^A}_B(x) \,.$ Thus, the subalgebra of gauge invariant
fields is, by definition, the commutant ${\cal A}({\cal G})' $ of
the algebra ${\cal A}({\cal G}) \, ,$ generated by the set $
\left\{ {{\cal G}^A}_B(x) \right\} \,.$

The local Gauss law at $x\in {\Lambda}^0$ has the form
\begin{equation}
\label{Gauss} \sum_y {E^A}_B(x,y) = {\rho^A}_B(x) \ ,
\end{equation}
with the sum taken over all points $y$ adjacent to $x \, .$
Imposing it on the subalgebra ${\cal A}({\cal G})' $ of gauge
invariant fields means factorizing the latter with respect to
${\cal I}(\Lambda ) \cap {\cal A}({\cal G})' \, ,$ where ${\cal
I}(\Lambda )$ is the ideal  generated by (\ref{Gauss}). Thus, the
observable algebra is defined as follows:
\begin{equation}
\label{observ} {\cal O}(\Lambda ) := {\cal A}({\cal G})' / \{{\cal
I}(\Lambda ) \cap {\cal A}({\cal G})'\} \, .
\end{equation}
Obviously, every element of ${\cal O}(\Lambda )$ is represented by
a gauge invariant element of ${\cal F}(\Lambda )$ and ${\cal
O}(\Lambda )$ can be viewed as a $*$-subalgebra of ${\cal
F}(\Lambda )$ generated by gauge invariant bosonic combinations of
$U$ and $E$ and by gauge invariant combinations of $\psi$ and
$\psi^*$ of mesonic and baryonic type, with the Gauss law inducing
some identities between those generators.

It will be shown in the next section that there is a (gauge
invariant), {\em additive} law, obtained by combining these two
equations, which characterizes the local colour charge density
carried by the lattice quantum fields. Additivity will allow us to
obtain the global colour charge by adding up the local Gauss laws.
In a subsequent paper we will show that the irreducible
representations of ${\cal O}(\Lambda )$ are labeled by this global
charge. This way, we get the physical Hilbert space ${\cal
H}^{phys}$ as a direct sum of colour charge superselection
sectors. The similar problem for QED on the lattice was solved in
\cite{KRT}.

\setcounter{equation}{0}
\section{The Local Charge density}
\label{Gauss Law}

In this section we are going to analyze the local Gauss law
(\ref{Gauss}). Suppose, for that purpose, that we are given a
collection of operators ${F^A}_B$ in a Hilbert space $\cal H$,
fulfilling ${F^A}_A = 0$ and $({F^A}_B)^*  =  {F_B}^A$, realizing
the canonical commutation relations for the Lie algebra $su(3)$:
\begin{equation}
\label{CCRF} [{F^A}_B , {F^C}_D] = {\delta^C}_B {F^A}_D  -
{\delta^A}_D {F^C}_B  \, .
\end{equation}

The field algebra  ${\cal F}(\Lambda )$ provides us with two basic
examples of this type: the electric field ${E^A}_B(x,y)$ on each
lattice link, see (\ref{CCR2}), and the charge operator
${\rho^A}_B(x)$ at each lattice site, given by formula
(\ref{rho}). Indeed, due to canonical anticommutation relations
(\ref{CCR3}), the latter fulfills (\ref{CCRF}). Thus, the
operators occurring on both sides of the local Gauss law fulfill
(\ref{CCRF}).

Throughout this paper, we assume integrability of the Lie algebra
representations under consideration. This means that for each $F$
there exists a unitary representation of the group $SU(3)$
\begin{equation}
SU(3) \ni g \rightarrow {\bar F}(g) \in B({\cal H}) \ ,
\end{equation}
associated with $F$.

It is easy to check that if $F$ and $G$ are two commuting
representations of $su(3)$ then also $F+G$ is. Indeed, if ${\bar
F}(g)$ and ${\bar G}(g)$ are representations of $SU(3)$
corresponding to $F$ and $G$, then $F+G$ may be obtained by
differentiating the representation $SU(3) \ni g \rightarrow {\bar
F}(g) {\bar G}(g)\in B({\cal H}) $, where $B({\cal H}) $ denotes
the $C^*$-algebra of bounded operators on ${\cal H}$. Moreover,
$-F^*$ is also a representation of $su(3)$, corresponding to the
following representation of $SU(3)$: $SU(3) \ni g \rightarrow
({\bar F}(g^{-1}))^* \in B({\cal H}) $.

Such a collection of operators is an {\em operator domain} in the
sense of Woronowicz (see \cite{kot}). We are going to construct an
operator function on this domain, i.~e.~a mapping $F \rightarrow
\varphi(F)$ which satisfies $\varphi(U F U^{-1} ) = U \varphi(F)
U^{-1}$ for an arbitrary isometry $U$. We are going to prove that
this function has the following properties:
\begin{eqnarray}
\varphi(- F^*) & = & - \varphi(F) \, , \label{fi2}\\ \varphi(F +
G) & = & \varphi(F) + \varphi(G) \label{fi3}\, ,
\end{eqnarray}
for commuting $F$ and $G$. This function will be built from the
two gauge-invariant, self-adjoint and commuting (Casimir)
operators $K_2$ and $K_3$ of $F$:
\begin{eqnarray}
K_2  & = & {F^A}_B {F^B}_A \, \\ K_3  & = &  \frac 12 \left(
{F^A}_B {F^B}_C {F^C}_A + {F^A}_B {F^C}_A {F^B}_C \right) \, .
\end{eqnarray}

The Hilbert space $\cal H$ splits into the direct sum of subspaces
${\cal H}_\alpha$ on which $F$ acts irreducibly. Each of these
subspaces is a common eigenspace of $K_2$ and $K_3$. Denoting the
highest weight characterizing a given irreducible representation
by $(m,n)$, with $m$ and $n$ being nonnegative integers, the
eigenvalues $k_2$ and $k_3$ of $K_2$ and $K_3$ are given by:
\begin{eqnarray}
k_2  & = &  \frac 23 (m^2 + mn + n^2 + 3m + 3 n) \, ,\label{k2}\\
k_3  & = &  \frac 19 (m - n)(3 + 2m + n)(3 + m + 2n) \, .
\label{k3}
\end{eqnarray}
It is easy to check that the above formulae may be uniquely solved
with respect to $m$ and $n$ yielding:
\begin{eqnarray}
m =M(k_2,k_3) & := &  \sqrt{ {\textstyle \frac 23} (k_2 + 2)}
\Bigg( \, \cos \left( \frac 13 \arccos{\frac{\sqrt{6} k_3}
{\sqrt{(k_2 + 2)^3}}} + \frac 23 \pi \right) + \nonumber\\ & & 2
\cos \left( \frac 13 \arccos{\frac{\sqrt{6} k_3}{\sqrt{(k_2 +
2)^3}}} \right) \Bigg) - 1 \, , \\ n =N(k_2,k_3) & := &  - \sqrt{
{\textstyle \frac 23} (k_2 + 2)} \Bigg( \, 2 \cos \left( \frac 13
\arccos{\frac{\sqrt{6} k_3}{\sqrt{(k_2 + 2)^3}}} + \frac 23 \pi
\right) + \nonumber\\ & & \cos \left( \frac 13
\arccos{\frac{\sqrt{6} k_3}{\sqrt{(k_2 + 2)^3}}} \right) \Bigg) -
1 \, .
\end{eqnarray}
Using these functions we may define a function with values in
${\mathbb Z}_3$, which may be identified with the center ${\cal
C}$ of the gauge group $SU(3)$:
\begin{equation}\label{modulo}
f(k_2,k_3) := (M(k_2,k_3) - N(k_2,k_3)) \, \, \mbox{\rm \bf mod}
\, 3 \, .
\end{equation}
For our purposes it is convenient to use the parametrization
${\mathbb Z}_3 = (-1 , 0 , 1)$. Since $K_2$ and $K_3$ are
commuting and self-adjoint, there exists an operator-valued
function:
\begin{equation}
\varphi(F) =  f(K_2(F),K_3(F)) \, .
\end{equation}
This means that $\varphi(F)$ may take eigenvalues $-1,0,1$ and
that every irreducible subspace ${\cal H}_\alpha$ is an eigenspace
of $\varphi(F)$ with eigenvalue $m - n$ {\bf mod} $\, 3$.

To prove property (\ref{fi2}) observe that  $K_2(-F^*) = K_2(F)$,
whereas $K_3(-F^*) = - K_3(F)$. Consequently, due to (\ref{k2})
and (\ref{k3}), we have $M(k_2,-k_3) = N(k_2,k_3)$ and
$N(k_2,-k_3) = M(k_2,k_3)$ which implies (\ref{fi2}). It remains
to prove property (\ref{fi3}). Let there be given two commuting
representations, $F$ and $G$, of $su(3)$ in $\cal H$. Denote the
irreducible components of $F$ by $\left\{ {\cal H}^F_{\alpha}
\right\}$ and of $G$ by $\left\{ {\cal H}^G_{\beta} \right\}$. The
irreducible spaces may be chosen in such a way that $\cal H$
decomposes as follows:
\begin{equation}
{\cal H} = \bigoplus_{\alpha , \beta} {\cal H}^F_{\alpha} \cap
{\cal H}^G_{\beta} \, .
\end{equation}
Take $0 \neq x \in {\cal H}^F_{\alpha} \cap {\cal H}^G_{\beta}$
and consider the space ${\cal H}^{F + G}_x \subset {\cal H}$
generated by vectors $\{ {\bar F}(g){\bar G}(g)x\}$, $g \in SU(3)
\, ,$ where ${\bar F}$ and ${\bar G}$ are the corresponding
(``integrated'') representations of $SU(3)$. By construction,
${\cal H}^{F + G}_x$ carries an irreducible representation of
${\bar F}{\bar G}\, .$ There exists a canonical embedding
\begin{equation}
{\cal H}^{F + G}_x \ni y \rightarrow T(y) \in {\cal H}^F_{\alpha}
\otimes {\cal H}^G_{\beta} \, ,
\end{equation}
given by
\begin{equation}
T \left( {\bar F}(g){\bar G}(g)x \right):= {\bar F}(g)x \otimes
{\bar G}(g) x\, ,
\end{equation}
intertwining the representation ${\bar F}{\bar G}$ with ${\bar
F}\otimes {\bar G}$. This means that ${\bar F}{\bar G}$, acting on
${\cal H}^{F + G}_x$, is equivalent to one of the irreducible
components of ${\bar F}\otimes {\bar G}$. Passing again to
representations of the Lie algebra $su(3)$, we conclude that $F +
G$, acting on ${\cal H}^{F + G}_x$, is equivalent to one of the
irreducible components of $F \otimes {\bf 1} + {\bf 1} \otimes G$,
acting on ${\cal H}^F_{\alpha} \otimes {\cal H}^G_{\beta}$. Now,
property (\ref{fi3}) follows from the following
\begin{lemma}
Let there be given two irreducible representations $(m,n)$ and
$(m',n')$ of $su(3)$, together with the decomposition of their
tensor product into irreducible components,
\begin{equation}
(m,n) \otimes (m',n') = (m_1,n_1) \oplus \dots \oplus (m_p,n_p) \,
.
\end{equation}
Then we have
\begin{equation}
(m - n)  \, \mbox{\rm \bf mod} \, 3 \, + (m' - n')  \, \mbox{\rm
\bf mod} \, 3 \, = (m_i - n_i)  \, \mbox{\rm \bf mod} \, 3 \,\, ,
\end{equation}
for every $i = 1, \dots , p\ $.
\end{lemma}

{\bf Proof:} We use the classification of irreducible
representations in terms of Young-tableaux. For $su(3)$, the
following two equations hold:
\begin{enumerate}
\item
The number of boxes constituting the Young-tableau of $(m_i,n_i)$
equals the sum of boxes of the tableaux corresponding to (m,n) and
(m',n') minus $3p$, where $p$ is a nonnegative integer.
\item
The number of boxes of an arbitrary irreducible representation of
$su(3)$ is equal to $m + 2n \equiv m - n + 3n \,$.
\end{enumerate}
Taking the first equation modulo three and using the second
equation yields the thesis.

Applying the operator function $\varphi$ to the local Gauss law
(\ref{Gauss}) and using additivity (\ref{fi3}) of $\varphi$ we
obtain:
\begin{equation}
\label{inGauss} \sum_y \varphi(E(x,y)) = \varphi(\rho(x)) \ .
\end{equation}
This is a gauge invariant equation for operators with eigenvalues
in ${\mathbb Z}_3$, valid at every lattice site $x$. The quantity
on the right hand side is the (gauge invariant) local colour
charge density carried by the quark field.

\setcounter{equation}{0}
\section{The Global Charge and the Flux Law}
\label{Charge}

Using the commutation relation between $E$ and $U$, transformation
law (\ref{ER-EL-1}) for $E(x,y)$  may be rewritten in three
equivalent ways:
\begin{eqnarray}\label{ER-EL-2}
   {E^A}_B(y,x) & = &
   {U^A}_D(y,x) {E^D}_C(x,y) {U^C}_B(x,y) + \frac 83 \delta^A_B \\
   & = & -
   {U^C}_B(x,y) {E^D}_C(x,y) {U^A}_D(y,x) - \frac 83 \delta^A_B \\
   & = & -
   {E^D}_C(x,y)  {U^A}_D(y,x) {U^C}_B(x,y) \ .\label{ER-EL-4}
\end{eqnarray}
These equations imply that
\begin{equation}\label{K2-K3}
  K_2(E(x,y)) = K_2(E(y,x)) \ , \ \ \ \ \ \ \ \ \
  K_3(E(x,y)) = - K_3(E(y,x)) \ .
\end{equation}
Hence, we have:
\begin{equation}
\varphi(E(x,y)) + \varphi(E(y,x)) = 0  \, ,
\end{equation}
for every lattice bond $(x,y)$.

Now we take the sum of equations (\ref{inGauss}) over all lattice
sites $x \in \Lambda$. Due to the above identity, all terms on the
left hand side cancel, except for contributions coming from the
boundary. This way we obtain the total flux through the boundary
$\partial \Lambda$ of $\Lambda$:
\begin{equation}\label{flux}
  \Phi_{\partial \Lambda} := \sum_{x \in \partial \Lambda}
  \varphi(E(x,\infty )) \ ,
\end{equation}
where by $E(x,\infty )$ we denote the colour electric charge along
the external link, connecting the point $x$ on the boundary of
$\Lambda$ with the ``rest of the world''. On the right hand side
we get the (gauge invariant) global colour charge, carried by the
matter field
\begin{equation}
\label{globalM} Q_{\Lambda} = \sum_{x \in \Lambda}
\varphi(\rho(x)) \, .
\end{equation}
Both quantities appearing in the global Gauss law
\begin{equation}\label{gG}
  \Phi_{\partial \Lambda} = Q_{\Lambda} \ ,
\end{equation}
take values in the center ${\mathbb Z}_3$ of $SU(3)$. The ``sum
modulo three'' is the composition law in ${\mathbb Z}_3$.

In the above discussion we have admitted non-zero values of $E(x,
\infty )$ at boundary points $x \in \partial \Lambda \, .$ In the
remainder of this section we make some remarks on the nature of
these objects. Our discussion will be rather heuristical, some
points will be made precise in a subsequent paper.
\begin{enumerate}
\item
One may treat $\Lambda$ as a piece of a bigger lattice
$\widetilde{\Lambda}$. Then the boundary flux operators $E(x,
\infty )$ belong to ${\cal F}(\widetilde{\Lambda})$ and commute
with ${\cal F}(\Lambda)$, (and also with ${\cal O}(\Lambda) \,) .$
They are external from the point of view of ${\cal F}(\Lambda)$
and measure the  ``violation of the local Gauss law'' on the
boundary $\partial \Lambda \, :$
\begin{equation}\label{ext}
E(x,\infty) := {\rho}(x) - \sum_y E(y,x)  \ .
\end{equation}
Non-vanishing of this element is equivalent to gauge dependence of
quantum states under the action of boundary gauges $g(x) \in SU(3)
\, , $ $x \in \partial \Lambda \, .$ Let us discuss this point in
more detail. Every irreducible representation of $SU(3)$ is
equivalent to some tensor representation. More precisely, denote
by $T^m_n({\mathbb C}^3)$ the space of $m$-contravariant,
$n$-covariant tensors over ${\mathbb C}^3$, endowed with the
natural scalar product induced by the scalar product on ${\mathbb
C}^3$. Let ${{\mathbb T}}^m_n({\mathbb C}^3) \subset
T^m_n({\mathbb C}^3)$ be the subspace of {\em irreducible},
i.~e.~completely symmetric and traceless tensors. These tensors
form a Hilbert space ${\cal T}({\mathbb C}^3)$ defined as the
direct sum
\begin{equation}
{\cal T}({\mathbb C}^3) := \bigoplus_{m,n} {{{\mathbb
T}}}^m_n({\mathbb C}^3) \, .
\end{equation}
Under gauge transformations at $x \in \partial\Lambda$, physical
states of QCD on $\Lambda$ behave like elements of ${\cal
T}({\mathbb C}^3)$, whereas the subspaces ${{\mathbb
T}}^m_n({\mathbb C}^3)$ correspond to eigenspaces of the invariant
operators $N(E(x,\infty))$ and $M(E(x,\infty))$, constructed from
external fluxes (\ref{ext}). If one wants to include all these
gauge invariant operators into an axiomatic formulation, as given
in Section \ref{Algebra}, one has different options. The remarks
at the beginning of this point suggest to postulate that these
operators commute with all elements of the observable algebra
${\cal O}(\Lambda )$. This corresponds to treating external fluxes
as purely {\em classical objects}, describing extra superselection
rules (cf.~\cite{Bu}, \cite{SW1}, \cite{S}).

\item
Our results concerning the charge superselection structure of QED
on a finite lattice \cite{KRT} suggest, however, a second option.
In \cite{KRT} all irreducible representations of the observable
algebra were classified in terms of the global electric charge
$Q_\Lambda$ contained in $\Lambda$. Representations differing only
by the local electric flux distribution over the boundary
$\partial \Lambda$, but having the same value of the global flux
$\Phi_{\partial \Lambda}$ were proved to be equivalent. The
redistribution of fluxes is obtained by the action of certain
unitary operators, see \cite{KRT}, acting on the quantum state
under consideration. Such a {\em redistribution operator} has the
following (heuristic) counterpart in continuum QED:
\begin{equation}\label{U-continuum}
  U(n) := \exp \left(\frac{i}{\hbar} \int_{\Sigma} n(x)
  \cdot A(x) \  d^3 x \right) \ ,
\end{equation}
where $n  = (n^{k})$ is a divergence-free (i.~e.~fulfilling
$\partial_k  n^{k} \equiv 0$) vector-density on $\Sigma \subset
{\mathbb R}^3$. Formally, we have:
\begin{equation}\label{E-continuum}
  {\tilde E}^k(x) :=  \left( U^*(n) E(x) U(n) \right)^k =
  E^k (x) + n^k (x) \ .
\end{equation}
It is obvious that replacing the field $E$ by ${\tilde E}$ and
leaving all other observables unchanged gives an equivalent
representation of the observable algebra. Nevertheless, the flux
field on the boundary $\partial \Sigma$ of the domain $\Sigma$ is
changed by $n^{\perp}(x)$, where ``$\perp$'' denotes the component
orthogonal to $\partial \Sigma$.

In a subsequent paper we are going to present a similar result for
lattice QCD. We shall prove that all irreducible representations
of the observable algebra ${\cal O}(\Lambda )$ of QCD on a finite
lattice are classified by the value of the global colour charge
$Q_{\Lambda} \, ,$ yielding three different superselection sectors
labeled by elements of ${\mathbb Z}_3 \, .$ However, the local
distribution of the (gauge invariant) gluon and anti-gluon fluxes
$M(E(x,\infty ))$ and $N(E(x,\infty))$ over the boundary $\partial
\Lambda$ may be arbitrarily changed within one sector. The
redistribution of fluxes is obtained by the following procedure.
Take an arbitrary pair of points $\xi , \eta \in \partial \Lambda$
at the boundary and a path (collection of lattice links) $\gamma =
\{ (\xi ,x_1), (x_1,x_2),\dots , (x_k,\eta )\}$, connecting them.
Define the following operator-valued matrix $U(\gamma) =
\left({U^A}_{B}(\gamma ) \right)$, where
\begin{eqnarray}\label{bigU}
  {U^A}_{B}(\gamma ) := \frac 1{\sqrt{3}} \  \ {U^A}_{C_1}(\xi ,x_1)
  {U^{C_1}}_{C_2}(x_1,x_2) \dots {U^{C_k}}_B(x_k,\eta ) \
   \ .
\end{eqnarray}
The action of $U(\gamma)$ on a quantum state $\psi$ is, by
definition, a collection $\left( {U^A}_{B}(\gamma ) \psi \right)$
with an extra contravariant index $A$ at $\xi$ and an extra
covariant index $B$ at $\eta$. In general, the new state does not
belong to any irreducible representation of $SU(3)$ at $\xi$ and
$\eta$, even if $\psi$ did. This means that $U(\gamma) \psi$ {\em
is not} an eigenstate of operators $M(E(\xi,\infty ))$,
$N(E(\xi,\infty ))$, $M(E(\eta,\infty ))$ and $N(E(\eta,\infty
))$, even if $\psi$ was. Decomposing it into irreducible
representations, we observe, however, that the value of
$\varphi(E(\xi,\infty ))$ has been changed by plus one and the
value of $\varphi(E(\eta,\infty ))$ has been changed by minus one
by this procedure. This suggests, that these objects should be
rather treated as {\em quantum} and not as classical quantities.
Only their sum, the global flux $\Phi_{\partial \Lambda}\, ,$ is a
classical object proportional to the identity on every
superselection sector. This point of view was strongly advocated
by Staruszkiewicz already a decade ago (see \cite{Star}). We also
refer to Giulini \cite{Giu}, who discussed decoherence phenomena
in QED in terms of {\em quantum} fluxes at infinity.

\item
Finally, we stress that it does not make sense to attribute any
physical meaning to the external gluon or anti-gluon fluxes $M$
and $N$ themselves. It is only the quantity $(M-N)$ ``modulo
three'' which makes sense. Heuristically, this again can be made
transparent for the continuum theory: Indeed, if we want to assign
the value of the flux through a piece $S \subset \partial \Sigma$
of the boundary of a domain $\Sigma$, we must be able to integrate
$E$ over $S$. Suppose, for that purpose, that $S$ has been divided
into small portions, $S = \bigcup_\alpha S_\alpha$. We have
\begin{equation}\label{sum}
  \int_S E = \sum_\alpha \int_{S_\alpha} E \ .
\end{equation}
The functions $M$, $N$ or even $(M-N)$ {\em are not} additive and,
therefore, cannot be applied to the left hand side in a way which
is compatible with the Riemann sums arising in the integration. On
the other hand, the function $\varphi$ is additive. This enables
us, in principle, to define the local ${\mathbb Z}_3$-valued flux
through $S$ as the sum of fluxes corresponding to its small
portions $S_\alpha$. Hence, if (one day in the future) continuum
QCD will be constructed as an appropriate limit of lattice
theories, also these surface fluxes will be defined as limits of
appropriate Riemann sums, corresponding to these lattice
approximations. We conclude that the only additive, gauge
invariant aspect of both the colour charge density $\rho$ and the
surface fluxes $E$ is carried by the ${\mathbb Z}_3$-valued
quantities $\varphi(\rho )$ and $\varphi(E)$.
\end{enumerate}

\section*{Acknowledgments}

The authors are very much indebted to J.~Dittmann, C.~\'Sliwa,
M.~Schmidt and I.P.~Volo\-bu\-ev for helpful discussions and
remarks. This work was supported in part by the Polish KBN Grant
Nr. 2 P03A 047 15. One of the authors (J.~K.) is grateful to
Professor E.~Zeidler for his hospitality at the Max Planck
Institute for Mathematics in the Sciences, Leipzig, Germany.


\end{document}